\documentclass[10pt]{article}\usepackage{amsmath,amssymb}
\usepackage[margin=1in]{geometry}

\newcommand{\vek}[1]{\mathbf{#1}} 
\newcommand{\unitvek}[1]{\mathbf{\hat{#1}}} 
\newcommand{\nn}[0]{\nonumber}
\newcommand{\lilbrk}[1]{\frac{_1}{^{#1}}}

\newcommand{\mat}[1]{\underline{\underline{\mathbf{#1}}}}
\newcommand{\mG}[0]{\mathrm{G}}
\newcommand{\mnKK}[0]{\mathrm{n}}
\newcommand{\bftaumat}[0]{\mat{\boldsymbol{\tau}}}

	\newcommand{\bfnabla}[0]{\boldsymbol{\nabla}}

\usepackage{graphicx}
\usepackage{dcolumn}
\usepackage{amsbsy}
\usepackage{multirow}
\usepackage[abs]{overpic}
\usepackage[numbers]{natbib}

\begin{document}
\title{Implementation of the Log-Conformation Formulation for Two-Dimensional Viscoelastic Flow}
\author{Kristian Ejlebjerg Jensen\\
kristianejlebjerg+cpc@gmail.com\\
Department of Micro- and Nanotechnology\\
 Technical University of Denmark \\
\vspace{5mm}\\
Fridolin Okkels\\
Department of Micro- and Nanotechnology\\
Technical University of Denmark  \\
\vspace{5mm}\\
Peter Szabo\\
ps@kt.dtu.dk\\
Department of Chemical and Biochemical Engineering\\
 Technical University of Denmark}



\maketitle

\section*{Abstract}
We have implemented the log-conformation method for two-dimensional viscoelastic flow in COMSOL, a commercial high-level finite element package. The code is verified for an Oldroyd-B fluid flowing past a confined cylinder. We are also able to describe the well-known bistability of the viscoelastic flow in a cross-slot geometry for a FENE-CR fluid, and we describe the changes required for performing simulations with the Phan-Thien-Tanner (PTT), Giesekus and FENE-P models. Finally, we calculate the flow of a FENE-CR fluid in a geometry with three in- and outlets. The implementation is included in the supplementary material, and we hope that it can inspire new as well as experienced researchers in the field of differential constitutive equations for viscoelastic flow.


\section{\label{sec:intro}Introduction}\vspace{-2pt}
The flow of viscoelastic fluids in complex geometries has many practical applications due to the fact that any dissolved elastic component, be it polymers or biological molecules, will give rise to viscoelastic effects. Such effects require the use of models that take the fluid's memory of past deformations into account. Differential constitutive equations is able to do just that, and they provide a good quantitative agreement with experiments \cite{baaijens1997viscoelastic}. Complex geometries call for the use of numerical methods, and therefore significant effort has been devoted towards various discretization techniques as well as model reformulations. The appearance of singularities, that caused codes to break down in smooth geometries at moderate elasticity, led to the definition of a ''High Weissenberg Number Problem'' (HWNP). This was effectively solved, when loss of positive definiteness for the conformation tensor was recognized as the cause of the singularity and the log-conformation method was introduced as a remedy \cite{hulsen2005flow}. 

Commercial numerical tools provide a simple workflow due to the integration of geometry description, unstructured mesh generation, discretization, solvers, and post-processing, which makes the treatment of complex geometries effortless compared to research grade code, but to our knowledge there does not exist any commercial tool that implements the log-conformation method. This is perhaps due to the fact that the method involves the calculation of eigenvectors and eigenvalues for the conformation tensor, which complicates the formulation of the constitutive equation. In two (and three) dimensions explicit expressions exists for these quantities, and it is thus possible to formulate the governing equations solely in terms of functions recognized by commercial simulation packages. We have included a COMSOL implementation of the Oldroyd-B model with log-conformation reformulation in the supplementary material, and we hope this can help other researchers working with differential constitutive equations. 

The article is structured in four parts: First we discuss implementation details of the log-conformation method for the most simple constitutive equations. We then calculate the flow of an Oldroyd-B fluid past a confined cylinder using COMSOL for comparison with other works \cite{hulsen2005flow}. As a third point we demonstrate the ability to predict bistable behavior of a FENE-CR fluid in a cross-slot geometry \cite{rocha2009extensibility}. Finally we consider a geometry with three in- and outlets, which has not previously been described in the context of viscoelastic flow, at least to our knowledge.

\section{Log-Conformation Implementation}\label{sec:impl}
\begin{figure}[!htb]
\centering \includegraphics[width= 0.16 \textwidth]{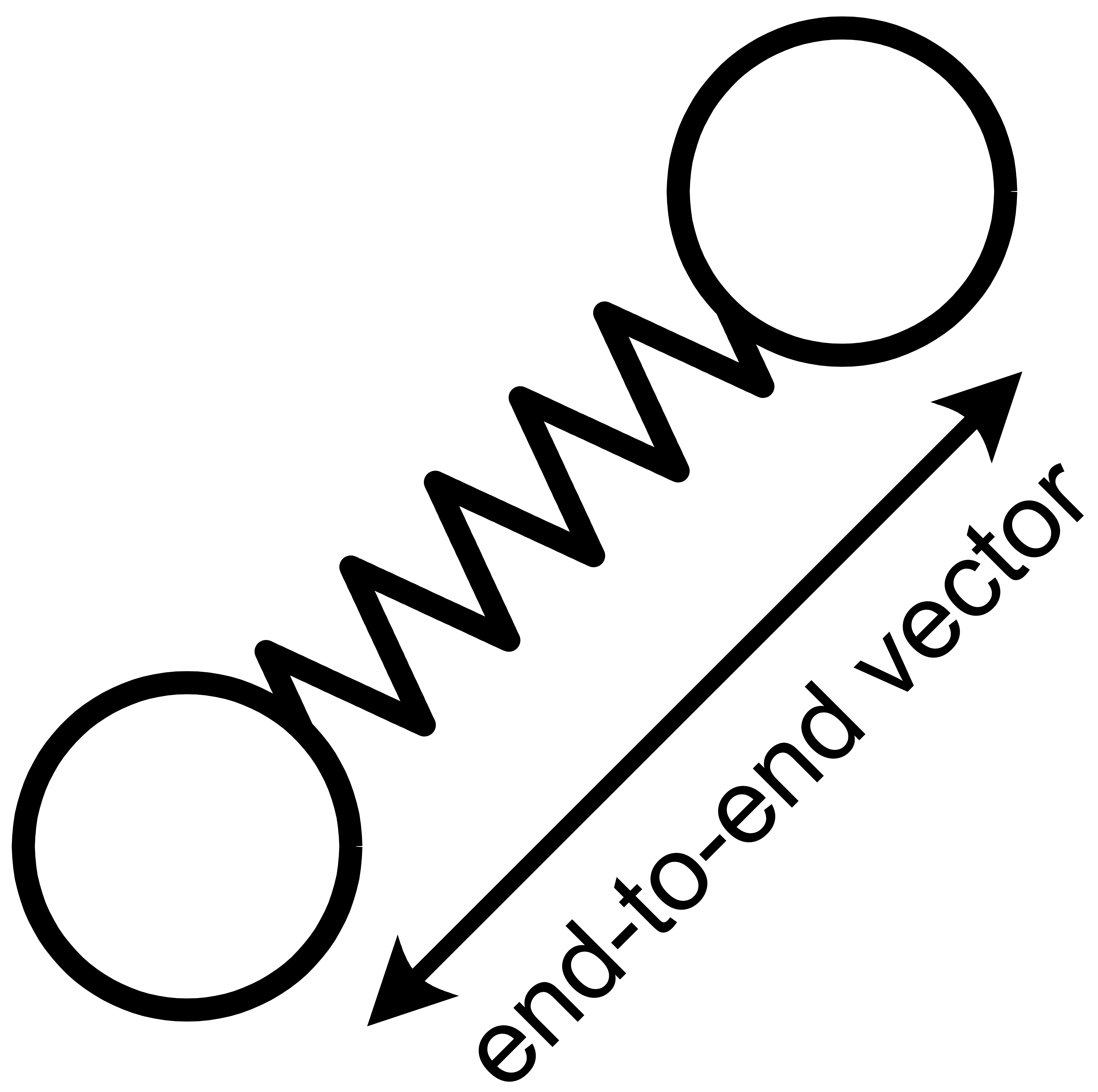}
\caption{An elastic dumbbell is illustrated with its two spring-connected point masses, and the end-to-end vector describing orientation and extension.} \label{fig:dumbbell}
\end{figure}

\vspace{-2pt}
In the so-called dumbbell models, the elastic part of a viscoelastic fluid is approximated by two spring-connected point masses -- an elastic dumbbell. The orientation and elongation of this dumbbell is described using the end-to-end vector $\vek{a}$ with the equilibrium length $\mathrm{a}_\mathrm{eq}$, see figure \ref{fig:dumbbell}. 
The conformation tensor, $\mat{A}$, is the statistical average, $\langle ... \rangle$ of the dyadic product between the end-to-end vector and itself normalized with the squared equilibrium length,
\begin{eqnarray}
\mat{A} &=& \frac{\left  \langle \vek{a}\otimes\vek{a} \right \rangle}{\mathrm{a}_\mathrm{eq}^2} . \label{eqn:a}
\end{eqnarray}
\vspace{-0.3cm}

The equation system for the Oldroyd-B model is written below. We neglect the effect of temperature and compressibility, but keep inertia for the sake of generality.
\begin{eqnarray}
\rho \frac{D\vek{v}}{Dt} &=& \vek{\bfnabla} \cdot \left(\overbrace{-p\mat{I} + \eta_s\left[\bfnabla\,\vek{v} + (\bfnabla\,\vek{v})^T \right]}^{\bftaumat_s} + \bftaumat_e\right) \label{eqn:navier} \\
\vek{\bfnabla} \cdot \vek{v} &=& 0 \label{eqn:cont} \\
\bftaumat_e &=& \frac{\eta_p}{\lambda}\left(\mat{A}-\mat{I}\right) \label{eqn:OBt} \\
0 &=& -\frac{1}{\lambda}(\mat{A}-\mat{I}) \nn \\
&+& \frac{D\mat{A}}{Dt}-\left[\mat{A}\cdot\bfnabla\,\vek{v}+(\bfnabla\, \vek{v})^T\cdot\mat{A}\right]\label{eqn:const} 
\end{eqnarray}
where $D$ is the material derivative, $\rho$ is the density, $\vek{v}$ is the velocity vector, $t$ is time, $p$ is the pressure $\eta_s$ is the solvent viscosity, $\eta_p$ is the elastic viscosity, $\lambda$ is the relaxation time, $\mat{\boldsymbol{\tau}}_s$ is the solvent stress tensor and $\mat{\boldsymbol{\tau}}_e$ is the elastic stress tensor. The Navier-Stokes equation (\ref{eqn:navier}) guarantees that Newton's 2nd law is obeyed, while the continuity equation (\ref{eqn:cont}) insures mass conservation. The expression for the elastic stress tensor in equation (\ref{eqn:OBt}) is best understood in light of the conformation tensor definition (\ref{eqn:a}) with $\eta_p/\lambda$ acting as a spring constant. The evolution of the conformation tensor is described in equation (\ref{eqn:const}) with relaxation on the first line balanced by the upper convected time derivative, which consists of convection in terms of the total derivative together with rotation and extension of the conformation tensor as imposed by the velocity gradient in square brackets.

When the velocity vector is approximated by $C^0$ continuous polynomials, the velocity gradient,  $\bfnabla\,\vek{v}$ becomes discontinous. This can cause numerical difficulties \cite{fortin2000discrete}, and therefore a $C^0$ continuous approximation $\mat{G}$ is usually constructed for use in equation (\ref{eqn:const}). Furthermore the zero, $\eta_p\left(\bfnabla\,\vek{v} + (\bfnabla\,\vek{v})^T - \mat{G} - \mat{G}^T\right)$ is added on the right-hand side of equation (\ref{eqn:navier}) to preserve the elliptic nature of the equation for small solvent viscosities in what is called the discrete elastic viscous stress splitting (DEVSS). Similarly, we use streamline upwind diffusion (SUPG) to introduce an elliptic component in the constitutive equation (\ref{eqn:const}), see section \ref{sec:Oweak}. 

The conformation tensor is symmetric, and the point of the log-conformation method is to exploit this property to rewrite equation (\ref{eqn:const}) to the form
\begin{eqnarray}
\frac{D\mat{s}}{Dt} - \Pi(\mat{s},\mat{G}) = 0, \nn
\end{eqnarray}
where $\mat{s}$ is the matrix logarithm of the conformation tensor. They are related by the following expression
\begin{eqnarray}
\mat{A} &=& \mat{R} \cdot e^{\mat{s}} \cdot \mat{R}^T \nn \\
\mat{R} &=& 
\left[\begin{array}{c c}
	v_{1\mnKK x} & -v_{1\mnKK y} \\
	v_{1\mnKK y} & v_{1\mnKK x} 
\end{array}\right]\nn \\
e^{\mat{s}} &=&
\left[\begin{array}{c c}
	e^{\lambda_1} & 0 \\
	0 & e^{\lambda_2}
\end{array}\right] \nn \\
A_{11} &=& v_{1\mnKK x}^2 e^{\lambda_1} + v_{1\mnKK y}^2 e^{\lambda_2} \nn \\
A_{12} &=& v_{1\mnKK x}v_{1\mnKK y} \left(e^{\lambda_1} - e^{\lambda_2} \right)\nn \\
A_{22} &=& v_{1\mnKK x}^2 e^{\lambda_2} + v_{1\mnKK y}^2 e^{\lambda_1} \nn 
\end{eqnarray}
where $\lambda_1$ and $\lambda_2$ are the eigenvalues of $\mat{s}$, while $v_{1\mnKK x}$ and $v_{1\mnKK y}$ are the normalized eigenvector components belonging to $\lambda_1$. The orthonormal rotation matrix, $\mat{R}$ holds the components of the eigenvectors, but we exploit that the second eigenvector is just the transpose of the first. Denoting the components of $\mat{s}$ by $s_{11}$, $s_{12}$ and $s_{22}$, the eigenvalues become\footnote{There also exist explicit expressions for the three-dimensional case.}
%
\begin{eqnarray}
\lambda_1 &=& \lilbrk{2}\left(s_{22}+s_{11}-\sqrt{(s_{22}-s_{11})^2+4s_{12}^2} \right) \quad \mathrm{and} \nn \\
\lambda_2 &=& \lilbrk{2}\left(s_{22}+s_{11}+\sqrt{(s_{22}-s_{11})^2+4s_{12}^2} \right) ,\nn
\end{eqnarray}
and the components of the first eigenvector are
\begin{eqnarray}
\vek{v}_1 &=& \left [ \begin{array}{c}
		\lambda_1-s_{22} \\
	s_{12}
\end{array} \right ] . \nn
\end{eqnarray}
The eigenvector is normalized
\begin{eqnarray}
\vek{v}_{1\mnKK} = \left[\begin{array}{c}
	v_{1x} \left/\sqrt{\vek{v}_{1x}^2+\vek{v}_{1y}^2} \right.\\
	v_{1y} \left/\sqrt{\vek{v}_{1x}^2+\vek{v}_{1y}^2} \right.
\end{array} \right] . \nn
\end{eqnarray}
The special case $s_{12}=0$ is handled by using $\mat{R}=\mat{I}$, when $|s_{12}|\leq\epsilon$ with $\epsilon = 10^{-12}$.

In the following we will express the reaction term $\mat{\Pi}$ as $\mat{R}\cdot\mat{\Omega}\cdot\mat{R}^T$ as described in \cite{kane2009comparison}. The continuous velocity gradient is rewritten in the principal frame
\begin{eqnarray}
\tilde{\mat{G}} &=& \mat{R}^T \cdot \mat{G} \cdot \mat{R} \nn \\
\tilde{\mG}_{11} &=& \vek{v}_{1nx}^2\mG_{11}+\vek{v}_{1nx}\vek{v}_{1yx}(\mG_{12}+\mG_{21}) + \vek{v}_{1ny}^2\mG_{22} \nn \\
\tilde{\mG}_{12} &=& \vek{v}_{1nx}^2\mG_{21}+\vek{v}_{1nx}\vek{v}_{1yx}(\mG_{22}-\mG_{11}) - \vek{v}_{1ny}^2\mG_{12} \nn \\
\tilde{\mG}_{21} &=& \vek{v}_{1nx}^2\mG_{12}+\vek{v}_{1nx}\vek{v}_{1yx}(\mG_{22}-\mG_{11}) - \vek{v}_{1ny}^2\mG_{21} \nn \\
\tilde{\mG}_{22} &=& \vek{v}_{1nx}^2\mG_{22}-\vek{v}_{1nx}\vek{v}_{1yx}(\mG_{12}-\mG_{21}) + \vek{v}_{1ny}^2\mG_{11} \nn.
\end{eqnarray}
Note that this matrix is not symmetric, and that we have chosen $\mG_{12}$ to refer to the derivative of the velocity vector $x$-component with respect to the $y$-coordinate.
The diagonal components $\Omega_{11}$ and $\Omega_{22}$ are 
\begin{eqnarray}
\Omega_{11} &=& 2 \tilde{\mG}_{11}  - \frac{1-e^{-\lambda_1}}{\lambda} \quad \mathrm{and} \nn \\
\Omega_{22} &=& 2 \tilde{\mG}_{22}  - \frac{1-e^{-\lambda_2}}{\lambda} . \nn
\end{eqnarray}
These expressions are specific to the Oldroyd-B model, but it is straight forward to write the expressions for the Giesekus, Phan-Thien-Tanner, FENE-P or FENE-CR models instead \cite{kane2009comparison}. 
\begin{eqnarray}
\mathrm{Giesekus}&:& \begin{array}{l l}\Omega_{11} =& 2 \tilde{\mG}_{11}  - f_\mathrm{Gie.}(\lambda_1) \\
\Omega_{22} =& 2 \tilde{\mG}_{22}  - f_\mathrm{Gie.}(\lambda_2) \end{array} \\ \nn
\mathrm{where} & & f_\mathrm{Gie.}(x) = \frac{1+\alpha_\mathrm{Gie.}(e^x-1)}{\lambda}(e^x-1) \nn \\
\mathrm{PTT(exp.)}&:& \begin{array}{l l}\Omega_{11} =& 2 \tilde{\mG}_{11}  - \frac{k_\mathrm{PTT}(\lambda_1,\lambda_2)}{\lambda} \nn \\
\Omega_{22} =& 2 \tilde{\mG}_{22}  - \frac{k_\mathrm{PTT}(\lambda_1,\lambda_2)}{\lambda} \nn \end{array}, \nn \\
\mathrm{where} & & k_\mathrm{PTT}(\lambda_1,\lambda_2)=e^{\epsilon_\mathrm{PTT}(\lambda_1+\lambda_2-3)} \\ \nn
\mathrm{FENE-P}&:& \begin{array}{l l}\Omega_{11} =& 2 \tilde{\mG}_{11}  - \frac{k(\lambda_1,\lambda_2)e^{\lambda_1}-1}{\lambda} \\
\Omega_{22} =& 2 \tilde{\mG}_{22}  - \frac{k(\lambda_1,\lambda_2)e^{\lambda_2}-1}{\lambda} \nn \end{array}, \nn \\
\mathrm{where} & & k(\lambda_1,\lambda_2)=\frac{1}{1-(\lambda_1+\lambda_2)/\mathrm{max}^2}\nn \\
\mathrm{FENE-CR}&:& \begin{array}{l l}\Omega_{11} =& 2 \tilde{\mG}_{11}  - k(\lambda_1,\lambda_2)\frac{e^{\lambda_1}-1}{\lambda} \\
\Omega_{22} =& 2 \tilde{\mG}_{22}  - k(\lambda_1,\lambda_2)\frac{e^{\lambda_2}-1}{\lambda} \nn \end{array}, \\ \nn
\mathrm{where} & & \quad k(\lambda_1,\lambda_2)=\frac{1}{1-(\lambda_1+\lambda_2)/a_\mathrm{max}^2}\nn
\end{eqnarray}
In the case of the FENE-P and FENE-CR models, the $a_\mathrm{max}$ is a maximum extension, i.e. $\mathrm{Tr}(\mat{A}) < a_\mathrm{max}^2$. The off-diagonal component, $\Omega_{12}$, equals
\begin{eqnarray}
\Omega_{12} = \left \{ \begin{array}{l l} \left(\frac{\lambda_1-\lambda_2}{e^{\lambda_1}-e^{\lambda_2}}\right)\left(e^{\lambda_1}\tilde{\mG}_{12}+e^{\lambda_2}\tilde{\mG}_{21}\right)&, |\lambda_1-\lambda_2|>\epsilon \\
\tilde{\mG}_{12}+\tilde{\mG}_{21} &, |\lambda_1-\lambda_2|\leq\epsilon \end{array}\right. \nn
\end{eqnarray}
where we have reused the numerical parameter for the eigenvectors to handle the special case of identical eigenvalues. We are now ready to compute the reaction term
\begin{eqnarray}
\Pi_{11} &=& v_{1\mnKK x}^2 \Omega_{11} - 2 v_{1\mnKK x}v_{1\mnKK y} \Omega_{12} + v_{1\mnKK y}^2 \Omega_{22} \nn \\
\Pi_{12} &=& \left(v_{1\mnKK x}^2 - v_{1\mnKK y}^2\right) \Omega_{12} + v_{1\mnKK x}v_{1\mnKK y} (\Omega_{11}-\Omega_{22}) \nn \\ 
\Pi_{22} &=& v_{1\mnKK x}^2 \Omega_{22} + 2 v_{1\mnKK x}v_{1\mnKK y} \Omega_{12} + v_{1\mnKK y}^2 \Omega_{11} \nn
\end{eqnarray}
To ease modification of the implementation, we have opted to construct functions for each of the above expressions instead of writing out $\mat{\Pi}$ in terms of expressions involving $\mat{s}$ and $\mat{G}$ only.

\paragraph{With respect to visualization} it is worth noting that $\mat{s}$ and $\mat{A}$ are positive definite tensors, and one can thus think of them as ''fields of ellipses''. In other words, the conformation tensor has an extension in every point given by its eigenvectors and eigenvalues. The trace of the conformation tensor is the sum of the extensions in the principal directions given by the eigenvectors, while the determinant is the product of these extensions. The trace and determinant equal the sum and product of the eigenvalues, respectively, and they are invariant with respect to rotation of the coordinate system, which make them interesting in the context of post-processing/visualization\footnote{Note that $\mathrm{det}(\mat{A})=\prod_ie^{\lambda_i}=e^{\sum{i}\lambda_i}=e^{\mathrm{Tr}(\mat{s})}$, where $\lambda_i$ are the eigenvalues of $\mat{s}$.}. The last invariant is the angle(s) related to the eigenvectors, but the eigenvector associated with the largest eigenvalue is normally oriented with the streamlines, so this angle provides little insight. Furthermore the trace and determinant often look very similar, at least on a logarithmic scale, and therefore we just plot the trace.

\subsection{Non-Dimensional Equations}
We define the dimensionless spatial coordinates, velocity, time, pressure and stress ($\tilde{x}$, $\tilde{v}$, $\tilde{t}$, $\tilde{p}$ and $\tilde{\bftaumat}_e$),
\begin{eqnarray}
\vek{x} = L_\mathrm{char} \vek{\tilde{x}}, \quad \vek{v} &=& v_\mathrm{char} \vek{\tilde{v}}, \quad t=\frac{L_\mathrm{char}}{v_\mathrm{char}}\tilde{t}, \nn \\
\quad p &=& \frac{v_\mathrm{char}(\eta_s+\eta_p)}{L_\mathrm{char}}\tilde{p} \nn \\
\mathrm{and} \quad \bftaumat_e &=& (\eta_s+\eta_p)\frac{v_\mathrm{char}}{L_\mathrm{char}} \tilde{\bftaumat}_e ,\nn 
\end{eqnarray}
which give rise to the following dimensionless constants
\begin{eqnarray}
\mathrm{Re} &=& \frac{\rho v_\mathrm{char} L_\mathrm{char}}{\eta}, \quad \mathrm{We} = \lambda \frac{v_\mathrm{char}}{L_\mathrm{char}} \quad \mathrm{and} \nn \\
\beta &=& \frac{\eta_s}{\eta_p+\eta_s} , \nn
\end{eqnarray}
The Weissenberg number, $\mathrm{We}$, describes the magnitude of elastic to viscous effects. When either the Weissenberg number goes to zero or the solvent to total viscosity ratio, $\beta$ goes to unity, the model approaches that of a Newtonian fluid. Contrarily, the value $\beta=0$ corresponds to the upper convected Maxwell model, which is characterized by particularly strong viscoelastic effects. Note that we prefer to define a characteristic pressure in terms of a characteristic velocity rather than the other way around due the fact that we intend to impose fixed flow rates rather than fixed driving pressures (as in \cite{ejlebjerg2012topology}).

We have implemented the above approach to the log-conformation formulation for two dimensions in COMSOL Multiphysics version 4.3a, and the online version of the manuscript includes a MATLAB script for computation of the flow of an Oldroyd-B fluid past a confined cylinder using the COMSOL LiveLink interface. We use default settings for everything, except we force update of the Jacobian every time a time step is saved. 

\subsection{The Oldroyd-B model in Weak Form} \label{sec:Oweak}
Converting the strong form of the Oldroyd-B model (\ref{eqn:navier}-\ref{eqn:const})	to weak form is trivial for the equations that relate to the elastic stress (\ref{eqn:OBt}) and continuity equation (\ref{eqn:cont}) as one just multiplies with test functions for the elastic stress and pressure, respectively. The evolution equation for the conformation tensor (\ref{eqn:const}) requires SUPG stabilization, which means that the equation is multiplied\footnote{Note that the test functions for elastic stress and conformation tensor are multiplied element wise (Hadamard product), denoted with $\circ$.} with $\mat{A}_\mathrm{test}+(2h_\mathrm{mesh}/v_\mathrm{char})\vek{v}\cdot\bfnabla \mat{A}_\mathrm{test}$ rather than $\mat{A}_\mathrm{test}$, where $h_\mathrm{mesh}$ is a characteristic size for the mesh. In other words a small amount of stream-wise diffusion is added to allow computation, but since the amount of diffusion is scaled with the mesh, the implementation still converges towards the solution without any artificial diffusion. Finally there is the momentum part of the Navier-Stokes equation (\ref{eqn:navier}), which is multiplied with a test function for the velocity and integrated by parts:
\begin{eqnarray}
0 &=& \left. \int_\Omega \left \{\tilde{\bfnabla} \cdot\left(\tilde{\bftaumat}_s + \tilde{\bftaumat}_e \right) - \mathrm{Re}\frac{D\vek{\tilde{v}}}{D \tilde{t}} \right \} \cdot \vek{\tilde{v}}_\mathrm{test} d \Omega \right. \nn \\
0 &=& -\left.\int_{\Omega} \tilde{\bftaumat}_s \circ \tilde{\bfnabla}\,\vek{\tilde{v}}_\mathrm{test}d\Omega\right. \\
&+& \left. \int_{\Omega} \left \{\tilde{\bfnabla}\tilde{\bftaumat}_e- \mathrm{Re}\frac{D\vek{\tilde{v}}}{D \tilde{t}} - \mathrm{Da}^{-1}\tilde{\alpha}\tilde{\vek{v}}\right\}\cdot \vek{\tilde{v}}_\mathrm{test} d\Omega \right. \nn \\
&+& \left.\int_{\partial \Omega} \tilde{\bftaumat}_s \cdot\unitvek{n}\cdot\vek{\tilde{v}}_\mathrm{test}dl \right.  \label{eqn:CRw}
\end{eqnarray}
The point is that the solvent stress is integrated by parts (Divergence theorem), but the elastic stress is not. This approach has been followed by other researchers so we have adopted it as well. The advantage is that the specification of $\tilde{\bftaumat}_s$ on open boundaries is slightly more simple than $\tilde{\bftaumat}_s+\tilde{\bftaumat}_e$, but in fact we have tested both without finding any difference. At outlets we thus usually impose normal velocity, zero pressure and 
\begin{eqnarray}
\tilde{\bftaumat}_s =  -\mat{I}\tilde{p} . \nn
\end{eqnarray}
This can also be used for pressure driven inlets, but in that case one should supplement the constraint on the velocity with one on the conformation tensor. We impose the no-slip($\vek{\tilde{v}}=\vek{0}$) at walls, which causes COMSOL to automatically enforce $\vek{\tilde{v}}_\mathrm{test}=0$, such that the boundary integral in equation (\ref{eqn:CRw}) vanishes.

\section{Confined Cylinder\label{sec:cyl}}
\begin{figure}[!htb]
\centering \includegraphics[width= 0.5 \textwidth]{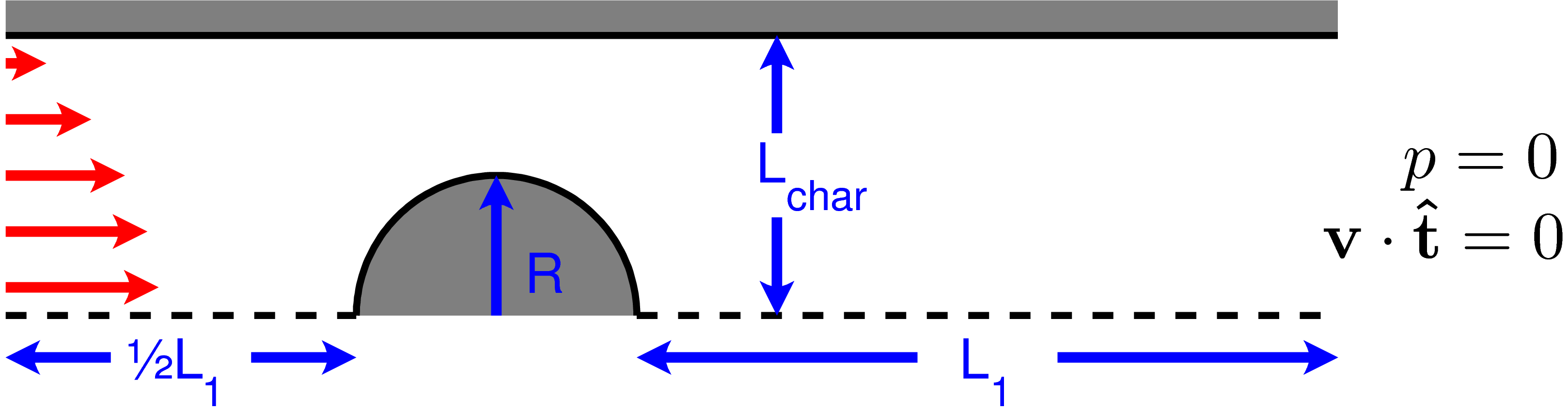}
\caption{The confined cylinder geometry with a (dashed) symmetry axis. The fluid comes in from the left, flows past the obstacle and exits through the boundary on the right ($\unitvek{t}$ is the unit tangent vector).} \label{fig:gcyl}
\end{figure}

The confined cylinder geometry is shown in figure \ref{fig:gcyl}, and it is a popular benchmark geometry due to the absence of geometric singularities. It can be used to verify an implementation by comparing the drag over the cylinder with a reference. We exploit the symmetry of the problem by modeling only half of the domain. 

The cylinder radius $R$, and average inlet velocity are used as characteristic length and velocity, respectively. The blocking ratio is set at 2 ($H=2R$), and the outlet length, $L_1$, is fixed at $10R$. 
%
In agreement with references \cite{hulsen2005flow} the solvent to total viscosity ratio is taken as 0.59, and inertia is neglected ($\mathrm{Re}=0$). The dimensionless drag is computed as
\begin{eqnarray}
\mathrm{Drag} &=& -2\int_\mathrm{cylinder} \boldsymbol{\tilde{\tau}} \cdot \unitvek{n} \cdot \unitvek{x}^T dl , \quad \mathrm{with} \nn \\
\mat{\boldsymbol{\tilde{\tau}}} &=& -\tilde{p}\mat{I} + \eta_s\left[\tilde{\bfnabla}\vek{\tilde{v}} + (\tilde{\bfnabla}\vek{\tilde{v}})^T \right] + \tilde{\bftaumat}_e \nn 
\end{eqnarray}
where $\unitvek{n}$ is the outward pointing normal vector.

We impose the no-slip boundary condition $\vek{\tilde{v}}=\vek{0}$ at the walls, and fully developed boundary conditions at the inlet
\begin{eqnarray}
\vek{\tilde{v}}_\mathrm{inlet} &=& \frac{_3}{^2}\left(1-(\tilde{y}/2)^2 \right)\unitvek{x} \nn \\  
A_\mathrm{11,inlet} &=& 1+2\left(\frac{\partial \left(\vek{\tilde{v}}\cdot\unitvek{x}\right)}{\partial \tilde{y}}\mathrm{We} \right)^2 = 1+\frac{_9}{^8} \tilde{y}^2\mathrm{We}^2  \nn \\ 
A_\mathrm{12,inlet} &=& \mathrm{We}\frac{\partial \left(\vek{\tilde{v}}\cdot\unitvek{x}\right)}{\partial \tilde{y}} = -\frac{_3}{^4} \tilde{y}\mathrm{We} \quad \mathrm{and} \\ \nn
A_\mathrm{22,inlet} &=& 1 \nn  
\end{eqnarray}
Note that the intermediate expressions can be used in the case of a pressure driven setup. At the outlet we impose zero pressure together with zero normal viscous stress.
\begin{eqnarray}
p_\mathrm{outlet} &=& 0 \quad \mathrm{and} \nn \\
\mat{\boldsymbol{\tilde{\tau}}}_{s,\mathrm{outlet}}  \cdot \unitvek{n} &=& \left(- \mat{I}p\right) \cdot \unitvek{n} = 0 \label{eqn:outlet}
\end{eqnarray}
At the symmetry line we impose zero normal velocity.
\begin{eqnarray}
\vek{\tilde{v}}_\mathrm{sym} \cdot \unitvek{n} = 0 \label{eqn:sym}
\end{eqnarray}

\subsection{Solution Details\label{ssec:prod}}
In order to find a steady solution one can go through the following steps.
\begin{itemize}
\item[\#1] Solve for the case of a Newtonian fluid.
\item[\#2] Use \#1 as initial condition for a transient simulation running for 10 relaxation times.
\item[\#3] Use \#2 as initial guess for a non-linear solver to find a steady solution.
\end{itemize}
We however choose to vary the Weissenberg number slowly using a regularized step function, $\mathrm{st}$, such that a quasi steady state is achieved.
\begin{eqnarray}
\mathrm{We} &=& \mathrm{We}_\mathrm{start} + \left(\mathrm{We}_\mathrm{end} -\mathrm{We}_\mathrm{start} \right)\mathrm{st}(\tilde{t}), \quad \mathrm{where} \nn \\
\mathrm{st}(\tilde{t}) &=& \left \{\begin{array}{l l l} 0 &,& \tilde{t} < \tilde{t}_\mathrm{start} \\
0.5+1.5\bar{t}-2\left(\bar{t}\,\right)^3 &, \,\tilde{t}_\mathrm{start} \leq & \tilde{t} < \tilde{t}_\mathrm{end} \\
1 &, \, \tilde{t}_\mathrm{end} \leq & \tilde{t} \end{array}\right. \nn \\
\bar{t} &=& \frac{\tilde{t}-(\tilde{t}_\mathrm{start}+\tilde{t}_\mathrm{end})/2}{\tilde{t}_\mathrm{end}-\tilde{t}_\mathrm{start}}\label{eqn:step}
\end{eqnarray}
This way we compute a quasi-steady solution to a range of Weissenberg numbers using a single transient simulation. COMSOL uses a fully implicit transient scheme with adaptive stepsize, so slow variation of the Weissenberg number does not increase the total computation time \cite{multiphysics20134}. 

\subsection{Results}
\begin{figure}[!htb]
\centering \includegraphics[width= 0.45 \textwidth]{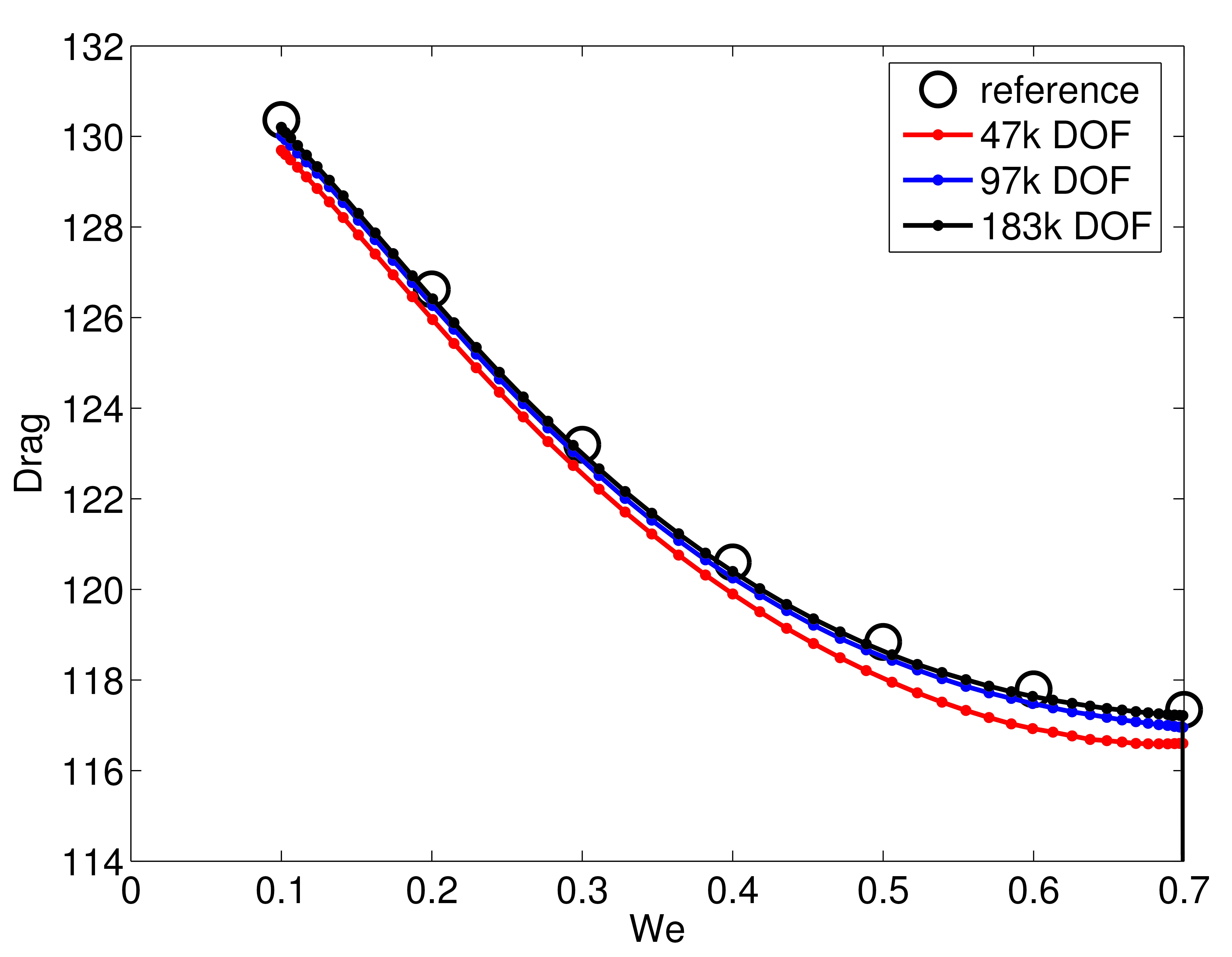}
\caption{The drag on a confined cylinder is plotted as a function of the Weissenberg number for three discretizations together with the data of another study \cite{hulsen2005flow}.} \label{fig:drag}
\end{figure}

Since the time over which the Weissenberg number is varied is a numerical parameter, we choose to scale it with the mesh size, so convergence towards the steady case is achieved, i.e.
\begin{eqnarray}
T_\mathrm{step} = \tilde{t}_\mathrm{end}-\tilde{t}_\mathrm{start} = 8000(0.07/h_\mathrm{mesh}) \nn
\end{eqnarray}
Figure \ref{fig:drag} shows the drag as a function of the Weissenberg number for three different discretizations indicating convergence up to $\mathrm{We}=0.6$. We observe good agreement with the reference drag although the simplicity of the chosen mesh gives rise to a high computational cost. It is well known, that convergence at $\mathrm{We}=0.7$ is difficult, even with anisotropic mesh adaption \cite{guenette2008adaptive}. In fact experimental \cite{mckinley1993wake} as well as theoretical evidence \cite{sahin2012parallel} exist to suggest, that a transition to a three dimensional flow pattern occurs, in other words the two dimensional symmetric flow becomes unstable. The three-dimensional solution has the flow alternating between going above and below the cylinder, and one can think of this solution as having a weaker extensional character in the wake of the cylinder.

\begin{figure}[!htb]
\centering \includegraphics[width= 0.45 \textwidth]{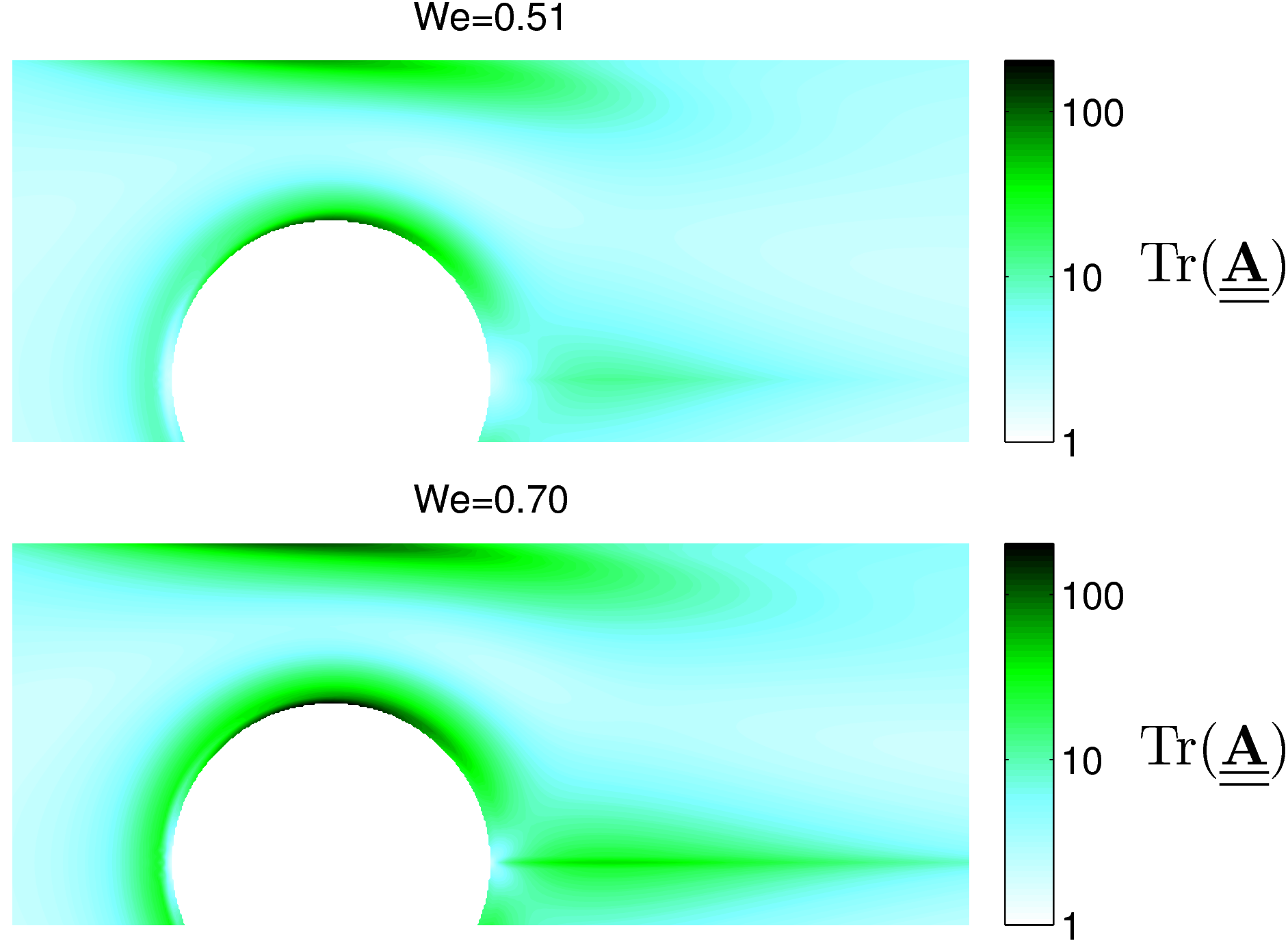}
\caption{The trace of the conformation tensor is plotted on a logarithmic scale for the confined cylinder in the case of $\mathrm{We}=0.51$ and $\mathrm{We}=0.7$ (183k DOF). The acceleration of the flow in the wake of the cylinder generates a strand of highly elongated dumbbells.} \label{fig:cylA}
\end{figure}
When a viscoelastic fluid enters a region, where the extension rate times the relaxation time 
reaches a critical value, the dumbbell extension starts to grow exponentially in time. This kind of non-linearity explains the significant difference between the wake of extension at $\mathrm{We}=0.51$ and $\mathrm{We}=0.7$ illustrated in figure \ref{fig:cylA}.

\section{Bistable Cross-Slot\label{sec:cslot}}
\begin{figure}[!htb]
\centering \includegraphics[width= 0.4 \textwidth]{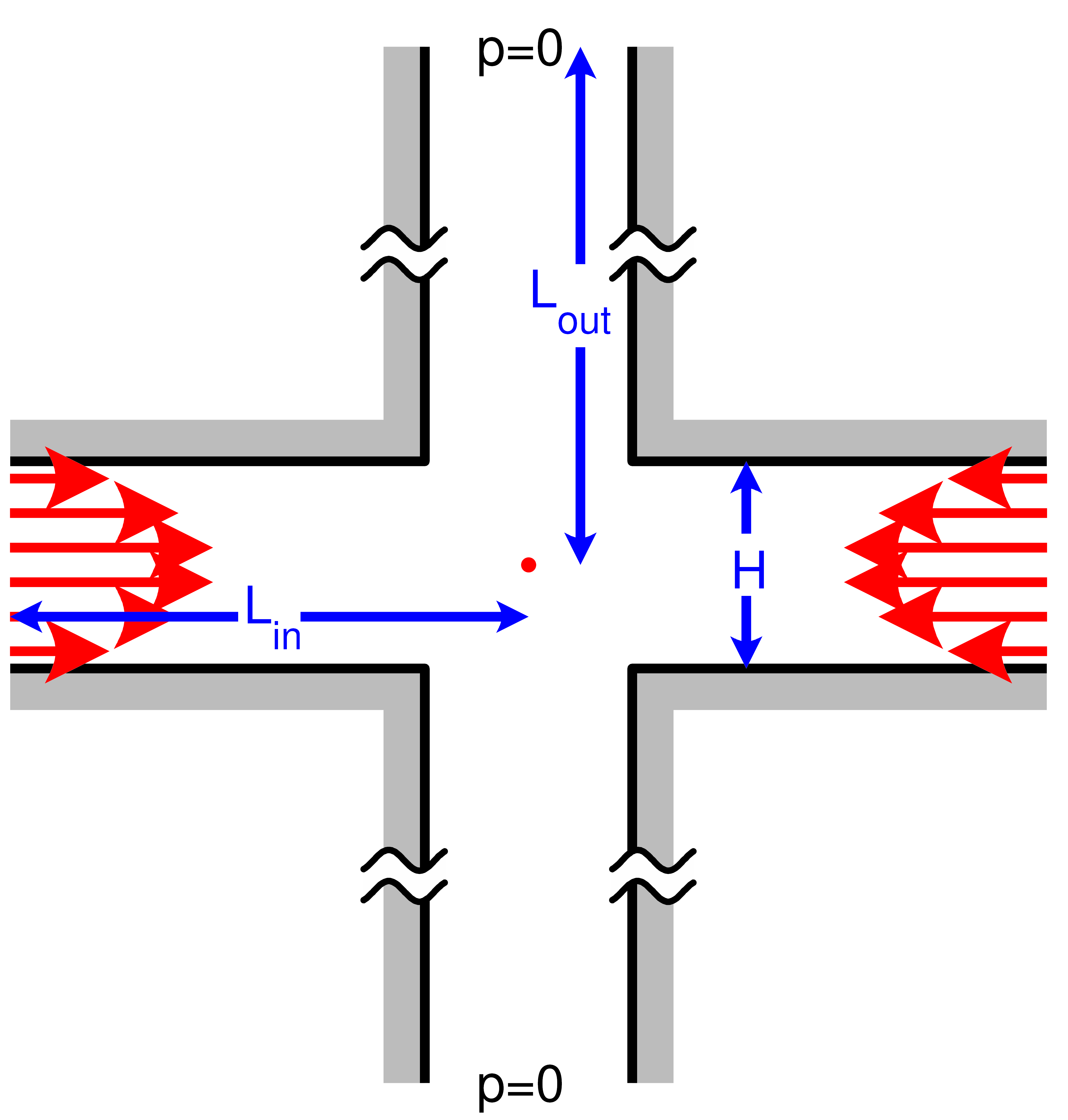}
\caption{The cross-slot geometry is shown with the fluid coming in from the left and right, diverging at the center, before exiting through to the upper and lower outlets leaving a free stagnation point (red) in the center of the geometry.} \label{fig:gcross}
\end{figure}

The FENE-CR model differs from the Oldroyd-B model in the sense, that $\lambda$ is replaced by $\lambda \left [1-\mathrm{Trace}(\mat{A})/a_\mathrm{max}^2 \right]$. In other words the relaxation time decreases as the dumbbell extension increases. 
This puts an upper bound $\mathrm{a}_\mathrm{max}^2$, on the trace of the conformation tensor corresponding to a maximum extension. The modification prevents an unbounded extension in free stagnation points such as the center of the cross-slot geometry shown in figure \ref{fig:gcross}. This geometry is known to exhibit bistability experimentally \cite{arratia2006elastic} as well as numerically \cite{poole2007purely}.

In this case we keep the average inlet velocity as  characteristic velocity and the channel width $H$ as characteristic length scale. We impose the no-slip boundary condition at walls, but we have been unable to initialize the transient simulation with the correct expressions for the conformation tensor at the inlets. Therefore we have used $\mat{s}=\epsilon\mat{I}$ at the inlets rather than values based on the expressions listed in Appendix I. Furthermore we had to change the damping strategy of the nonlinear method from the default (constant) to automatic.

\begin{figure}[!htb]
\centering \includegraphics[width= 0.45 \textwidth]{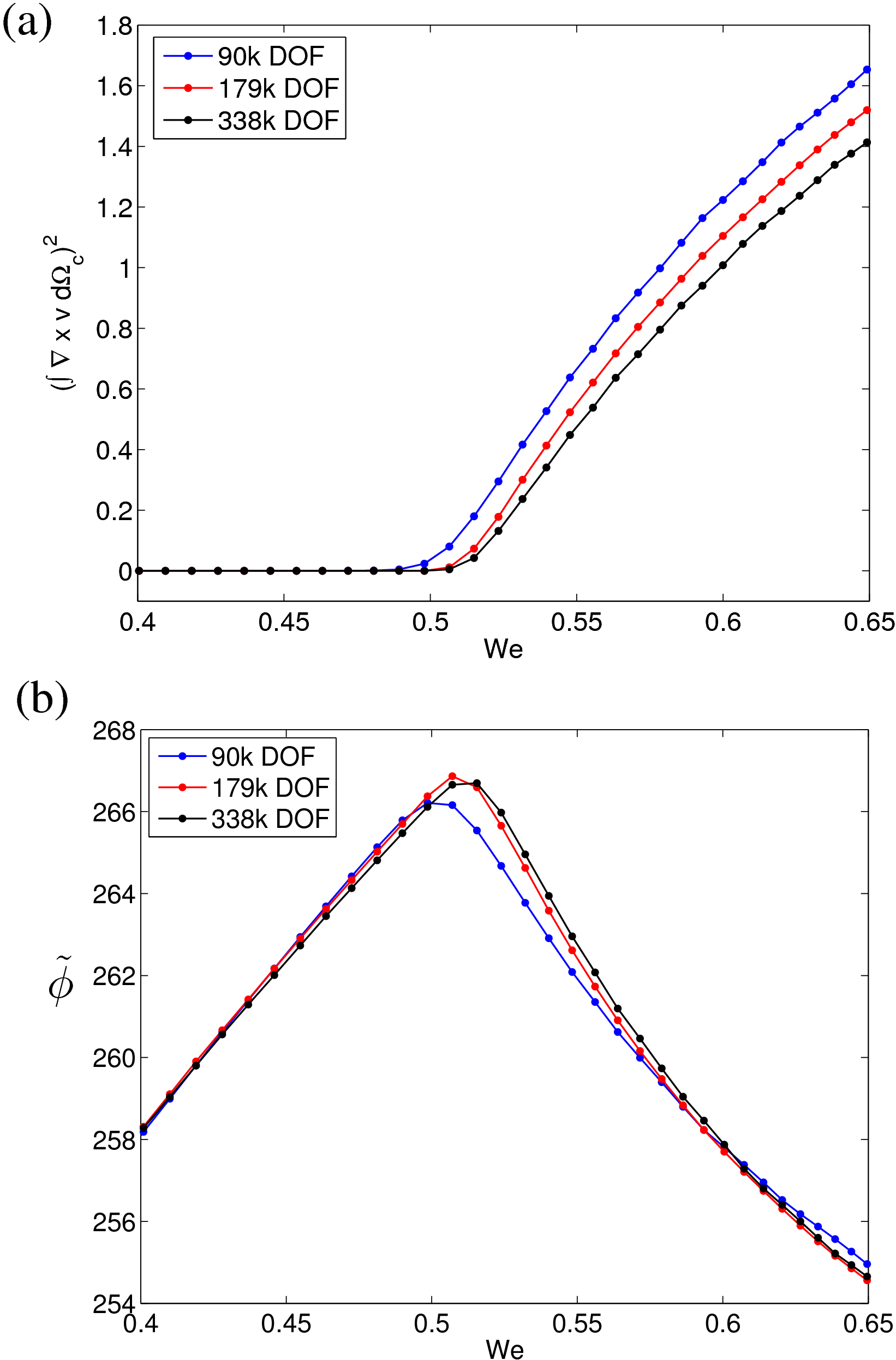}
\caption{The square of the vorticity integrated over the center of the domain $\Omega_c$ is plotted as function of the Weissenberg number in (a). In (b) it is the dissipation that is plotted, which shows that the point of bistability occurs ($\mathrm{We}=0.5$) at the point of maximum dissipation and thus also maximum hydraulic resistance.} \label{fig:flip}
\end{figure}

\begin{figure}[!htb]
\centering \includegraphics[width= 0.45 \textwidth]{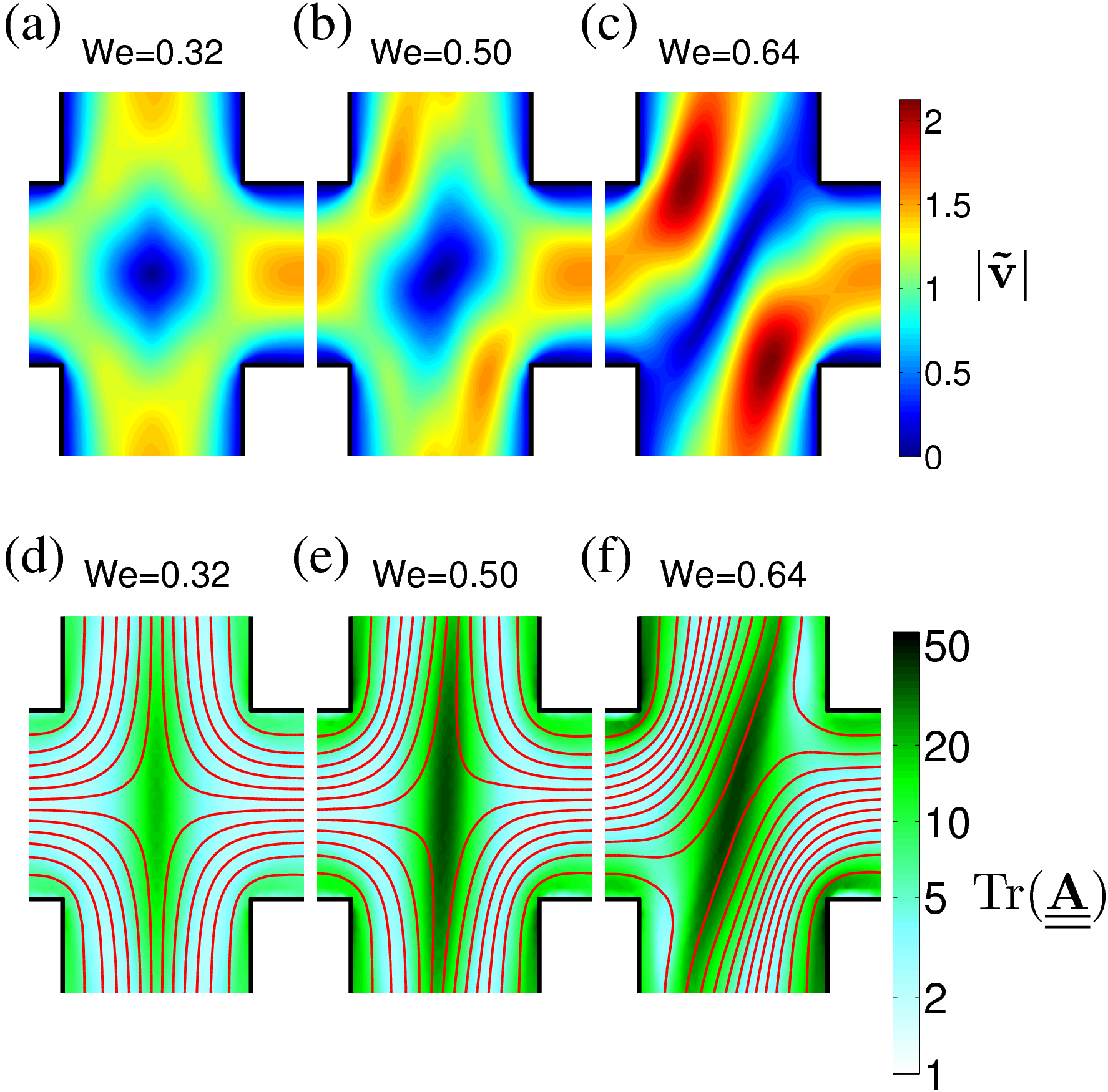}
\caption{The magnitude of the velocity vector is plotted for three Weissenberg numbers in the cross geometry in (a-c), while red stream lines are plotted on top of the trace of the conformation tensor on a logarithmic scale in (d-f). Note that we plot streamlines as contours of the stream function instead of integrating the velocity field.} \label{fig:cross}
\end{figure}

At the outlet we impose the same boundary conditions (\ref{eqn:outlet}) as for the cylinder. We compute the dissipation as an integral over the entire computation domain,
\begin{eqnarray}
\tilde{\phi} = \int_\Omega \mat{\boldsymbol{\tilde{\tau}}} : \left[ \tilde{\bfnabla}\vek{\tilde{v}} + (\tilde{\bfnabla}\vek{\tilde{v}})^T \right] d\Omega . \label{eqn:phi}
\end{eqnarray}
where $:$ is the Frobenius product. We choose $\mathrm{Re}=0$, $\mathrm{a}_\mathrm{max}^2=100$ and $\beta=0.2$ and the same transient solution procedure as described in section \ref{ssec:prod}. We define the center of the domain, 
\begin{eqnarray}
\Omega_c&=&\vek{\tilde{x}}\in|\vek{\tilde{x}}|_\infty<0.5, \nn
\end{eqnarray}
as we find that the vorticity integrated over this is a good asymmetry parameter. We thus plot the square of this versus the Weissenberg number in figure \ref{fig:flip}(a). The system becomes bistable at $\mathrm{We}=0.5$\footnote{Based on personal correspondence we have concluded that the results of \cite{rocha2009extensibility} are based on the FENE-MCR model, so no reference exists for the FENE-CR model.}, which gives rise to a kink in the dissipation as shown in figure \ref{fig:flip}(b). The solution before the kink [figure \ref{fig:cross}(a) and (d)] has vertical, horizontal as well as 180-degree rotational symmetry, while the solutions after the kink [\ref{fig:cross}(b), (c), (e) and (f)] only have 180-degree rotational symmetry, and we will refer to these solutions with reduced symmetry as ''asymmetric solutions''. The phenomenon can be understood by considering the velocity magnitude and conformation tensor trace as plotted in figures \ref{fig:cross}(a-c) and (d-e) respectively. The dumbbells are extended along the vertical axis, which causes a damping that delays the merging of the flows, and this gives rise to extra shear and thus also dissipation. When the solution becomes asymmetric, the damping is moved to the side of the channel, which lowers the dissipation. One can also think of the asymmetry as a flow pattern with less extension in the stagnation point, similar to the three-dimensional flow for the confined cylinder.

\section{Three Inlets}
\begin{figure}[!htb]
\centering \includegraphics[width= 0.4 \textwidth]{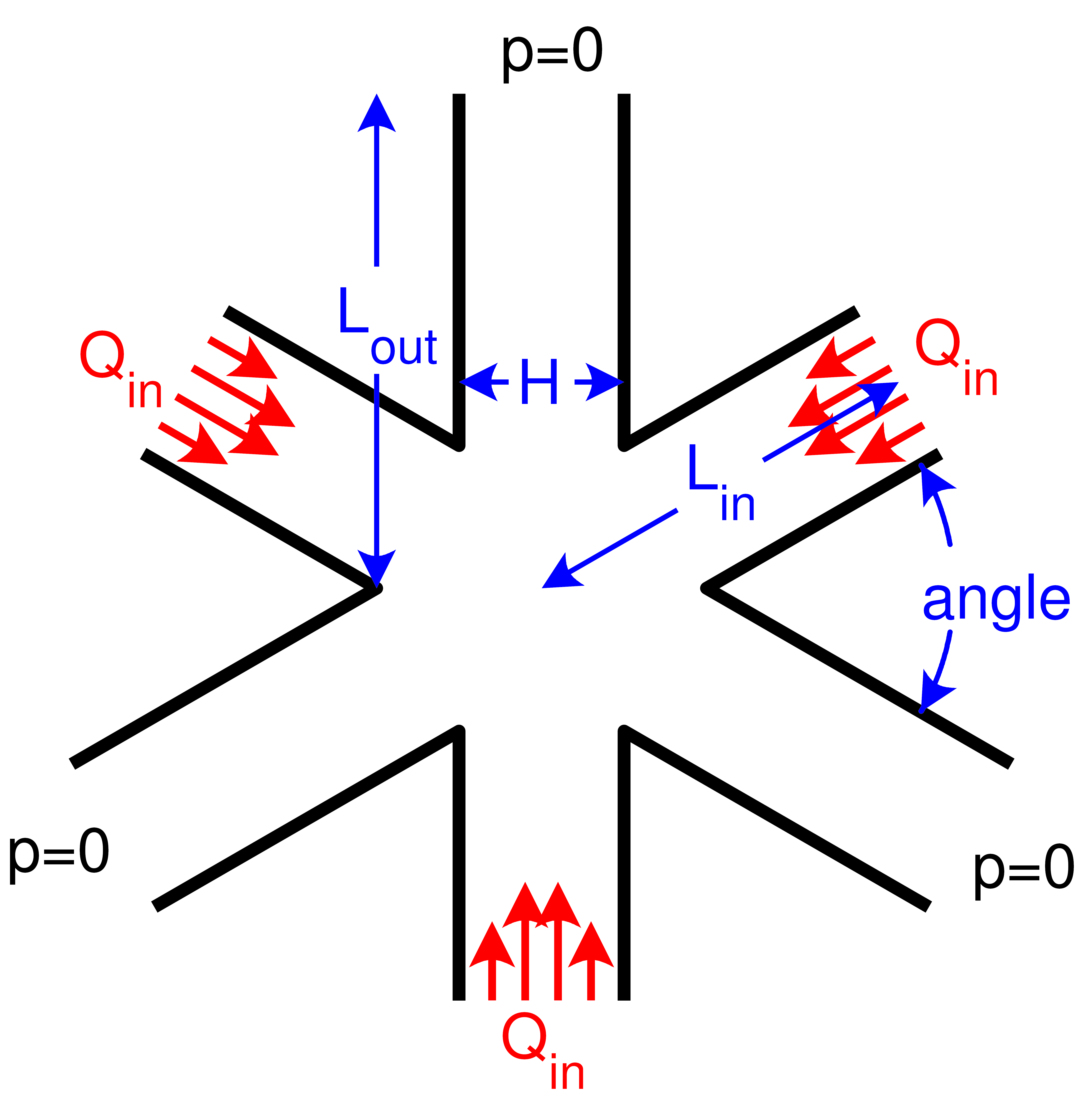}
\caption{The geometry with three in- and outlets is illustrated with the different length scales and angles involved. For an angle of $\pi/3$ the geometry has a symmetry axis through every channel as well as 120-degree rotational symmetry, while an angle different from $\pi/3$ breaks all symmetries except the one around the vertical axis.} \label{fig:sgeom}
\end{figure}

In this section we introduce the geometry shown in figure \ref{fig:sgeom}. It consists of a hexagon with in- and outlets attached in an alternating fashion. The geometry always has a vertical symmetry axis, but depending on whether a certain angle is equal to $\pi/3$ or not, it can also have 120-degree rotational symmetry. We define the distance to the center as in- and outlet lengths, and set equal inlet flow rates as well as outlet pressures. We set $L_\mathrm{in}=6H$ and $L_\mathrm{out}=8L$, $\beta=0.1$ and $\mat{s} = \epsilon\mat{I}$ at the inlets, but keep all other parameters identical to those used for the cross-slot. That is, the channel width and average inlet velocity are still used as characteristic length and velocity, respectively.

\begin{figure}[!htb]
\centering \includegraphics[width= 0.45 \textwidth]{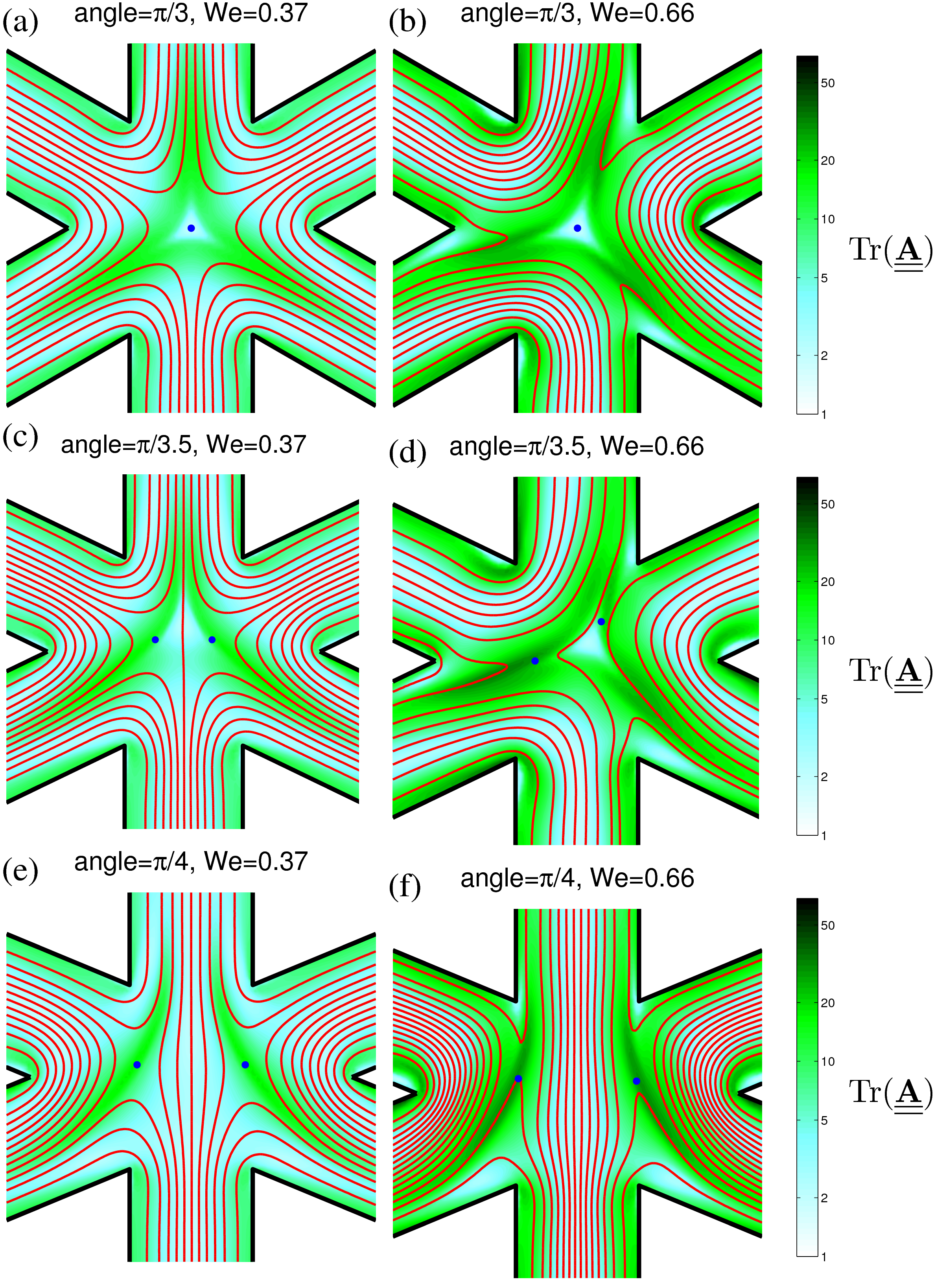}
\caption{The trace of the conformation tensor is plotted on a logarithmic scale together with streamlines (red) for two Weissenberg numbers in a geometry with three inlets. The case of an angle equal to $\pi/3$ (a-b), $\pi/3.5$ (c-d) and $\pi/4$ (d-e) radians is illustrated (see figure \ref{fig:sgeom}). Although not apparent from the streamlines, the upper outlet receives fluid from all three inlets in (d), and the flow between the stagnation points is reduced with a factor of 4.5 in (d) compared to (c). Any asymmetry around the vertical axis in (a), (c) and (e) is due to the mesh.} \label{fig:sres}
\end{figure}

If the angle is equal to $\pi/3$ radians, a single stagnation point exists in the center of the geometry regardless of whether we consider Stokes flow or just the regime of low elasticity. Two stagnation points however appear, if the angle is different from $\pi/3$ radians. Figure \ref{fig:sres} shows the solution for angles of $\pi/3$, $\pi/3.5$ and $\pi/4$ radians at $\mathrm{We}=0.37$ (low) and $\mathrm{We}=0.66$ (moderate). For an angle of $\pi/3$ radians, the regime of low elasticity (a) gives a solution with 120-degree rotational- and vertical symmetry, while for moderate elasticity a solution with only 120-degree rotation symmetry appears (b). This latter solution involves the clockwise flow, but by symmetry we can argue that there must be an identical solution with counter-clockwise flow. The symmetry reduction also occurs for an angle of $\pi/3.5$ (c-d), in the sense that we go from only vertical symmetry to no symmetry. Regardless of the Weissenberg number, there are two stagnations points even though the flow rate between them is small at moderate elasticity (d). In other words turning up the elasticity causes the flow rate between the stagnation points to decrease significantly. As shown in (e-f) this is different for an angle of $\pi/4$ radians, as the elasticity tends to move the stagnation points apart and increase the flow rate between them in what we call the ''straight flow solution'', that is the solution has identical symmetry properties at low and moderate elasticity.

%

\begin{figure}[!htb]
\centering \includegraphics[width= 0.45 \textwidth]{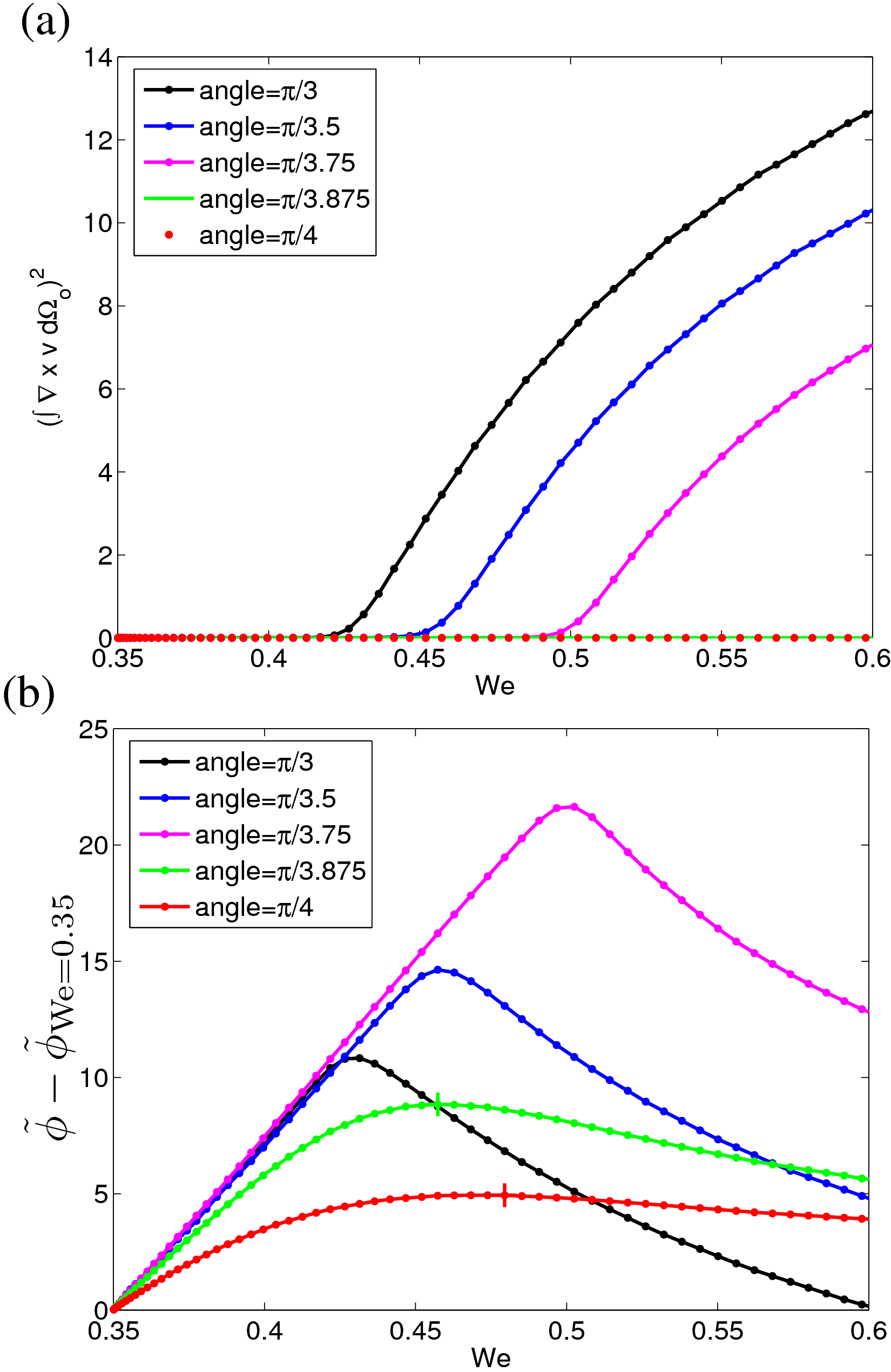}
\caption{The square of the vorticity integrated over the center of the domain $\Omega_o$ is plotted as functions of the Weissenberg number for five different angles in (a), while the dissipation is shown (b). The point of bistability coincides with the point of maximum dissipation for the three larger angles, as it was also the case for the cross-slot in figure \ref{fig:flip}. The maximum dissipation of the two smaller angles is indicated with crosses.} \label{fig:s}
\end{figure}

Using artificial initial conditions as listed in Appendix II, one can show that the clockwise, counter-clockwise and the straight flow all exist as stable solutions at moderate elasticity for angles between $\pi/4$ and $\pi/3.5$. This means, that it is the connectivity with the solution at low elasticity that changes. Furthermore there is an unstable solution with a vertical symmetry axis for the angles $\pi/3$ and $\pi/3.5$. We calculate the integral of the vorticity around the center 
\begin{eqnarray}
\Omega_o&=&\vek{x}\in||x||_2<0.5, \nn
\end{eqnarray}
as a measure of asymmetry and plot this as a function of the Weissenberg number for five angles in figure \ref{fig:s} (a), while the dissipation is plotted in figure \ref{fig:s} (b). It is clear that the point of maximum dissipation coincides with the point of bistability for the three largest angles ($\pi/3$, $\pi/3.5$, $\pi/3.75$), as it was also the case for the cross-slot in section \ref{sec:cslot}. The dissipation for the two smaller angles also have maxima, but we do not believe they play the same role as the maxima for the smaller angles. This is due to the fact that the Weissenberg number at which the maximum occurs seems to decreases for both the rotation and straight flow solutions. This means that there does not exists an angle for which the two kind of maxima in figure \ref{fig:s} (b) meet, although there must exist an angle where all three solutions are connected to the solution at low elasticity.

\section{Conclusion}
We have described the viscoelastic Oldroyd-B model with log-conformation reformulation in detail, and included a COMSOL implementation in the supplementary material. We have showed, that the code agrees with a reference in a benchmark geometry. Furthermore we have demonstrated that the code can be modified to describe the flow of a FENE-CR fluid, and that it can reproduce bistability in the cross-slot geometry. Finally we computed the flow in a geometry with three in- and outlets, which appears to feature three stable solutions. The connectivity of the solutions with the solution at low elasticity seems to depend on an angle in the geometry.

\providecommand{\noopsort}[1]{}\providecommand{\singleletter}[1]{#1}%

\section*{Appendix I: FENE-CR Inlet Conditions}
The developed flow conditions for a FENE-CR fluid \footnote{Note that one has to set $A_\mathrm{11,inlet}=1$ when $\chi^2<\epsilon$.} is listed in the following equations
\begin{eqnarray}
\vek{\tilde{v}}_\mathrm{inlet} &=& -\frac{_3}{^2}\left(1-(\tilde{y}/2)^2 \right)\unitvek{n} \nn \\  
A_\mathrm{11,inlet} &=& a_\mathrm{max}^2-1 \nn \\
&+&\frac{a_\mathrm{max}^2\left(a_\mathrm{max}^2-\sqrt{a_\mathrm{max}^4+8\chi^2(a_\mathrm{max}^2-2)}\right)}{4\chi^2} \nn \\ 
A_\mathrm{12,inlet} &=& \chi \left(1-\frac{A_\mathrm{11,inlet}+1}{a_\mathrm{max}^2}\right) \nn \\
A_\mathrm{22,inlet} &=& 1 , \quad \mathrm{where} \nn \\
\quad \chi &=& \mathrm{We}\frac{\partial \left(\vek{\tilde{v}}\cdot\unitvek{x}\right)}{\partial \tilde{y}} = 3\mathrm{We}(\tilde{y}/4)(\unitvek{n}\cdot\unitvek{x})	\nn  
\end{eqnarray}

\section*{Appendix II: Artificial Initial Conditions for the Geometry with Three In- and Outlets}
One can find stable solutions not connected to the solution at low elasticity using artificial initial conditions. A rotating solution for an angle of $\pi/4$ (see figure \ref{fig:sgeom}) can in example be found by imposing a rotating force in the center of the geometry,

\begin{eqnarray}
\vek{F}^{\circlearrowright} = \left \{ \begin{array}{l l l l l} (\unitvek{x}\tilde{y}-\unitvek{y}\tilde{x})/2 &,& ||\tilde{\vek{r}}||  & \leq & 0.5/\sin(\pi/6)\\
0 &,& ||\tilde{\vek{r}}|| & > & 0.5/\sin(\pi/6) \end{array}\right. \nn ,
\end{eqnarray}
where $\tilde{\vek{r}}$ is the position vector and  $||\cdots||$ is the euclidean norm. This is then multiplied with a step function as given by equation (\ref{eqn:step}), such that the force vanishes beyond some critical time. Alternatively one might find a solution going straight through a geometry with an angle of $\pi/3.5$ using an upward force,
\begin{eqnarray}
\vek{F}^{\uparrow} = \left \{ \begin{array}{l l l l l} \unitvek{y}(0.25-\tilde{x}^2) &,& ||\tilde{\vek{r}}|| & \leq & 0.5/\sin(\pi/6)\\
0 &,& ||\tilde{\vek{r}}|| & > & 0.5/\sin(\pi/6) \end{array}\right. \nn .
\end{eqnarray}
\end{document}